# A New Paradigm for Reconfigurable Intelligent Surface Design: Multi-port Network Method

Zhen Zhang, *Member, IEEE*, Qiang Cheng, *Senior Member, IEEE*

*Abstract*—As a novel approach to flexibly adjust the wireless environment, reconfigurable intelligent surfaces (RIS) have shown significant application potential across various domains, including wireless communication, radar detection, and the Internet of Things. Currently, mainstream design methods for reconfigurable intelligent surfaces face inherent limitations. For instance, while the full-wave electromagnetic (EM) simulation method offers strong universality, it suffers from low efficiency. Machine learning-based methods can effectively reduce design time but are heavily dependent on full-wave EM simulations. Although the design methods based on the equivalent circuit can lessen the reliance on full-wave EM simulations, they still struggle with insufficient model accuracy when dealing with complex element structures. In recent years, a new multi-port network method has been introduced to RIS design. This method has significantly enhanced the accuracy of modeling complex structures. It reduces the dependency on full-wave EM simulations and substantially shortens the design time. This work provides a detailed exploration of the RIS element design strategy based on multi-port network and discusses future development trends in this field.

*Index Terms*—Multi-port network, reconfigurable intelligent surface

## I. Introduction

Reconfigurable Intelligent Surfaces (RISs) possess the capability of flexibly reconfiguring the electromagnetic (EM) environment and can be applied in various fields such as wireless communication, radar, and wireless sensing [1]-[3]. For instance, with the rapid development of 5G technologies, spectral resources are becoming increasingly scarce. RISs can achieve efficient communication among multiple users in the same frequency band by precisely manipulating EM waves. They can also construct additional signal transmission links to solve the problem of signal obstruction between urban buildings, ensuring smooth indoor and outdoor communication and reducing the construction cost of base stations [4]-[6]. In the field of radar detection, RISs can dynamically alter the Radar Cross Section (RCS) characteristics of targets to evade enemy detection [7]-[10]. Regarding the Internet of Things, RISs can optimize the connection of low- power-consumption devices. By virtue of their own EM sensing ability, they can make device deployment more convenient in scenarios such as smart homes and smart agriculture and ensure more stable operation [11]-[14].

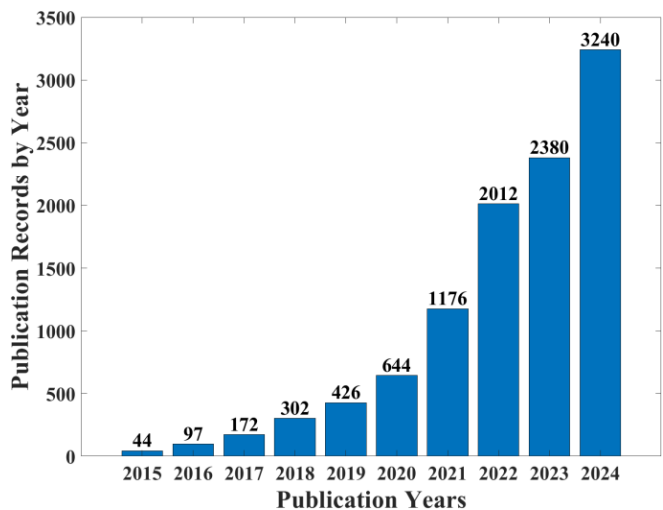

Fig. 1. Number of published literatures on reconfigurable intelligent surface-related topics from 2015 to 2024.

Based on the Web of Science database, we conducted a literature search for the period from 2015 to 2024. The search covered topics including "Reconfigurable Metasurface", "Reconfigurable Intelligent Surface", "Programmable Metasurface", "Information Metasurface", and "Coding Metasurface". As a result, we obtained a total of 10,493 relevant publications. Through the statistical analysis of the literature (see Fig. 1), it can be observed that over the past decade, the enthusiasm for research in the field of Reconfigurable Intelligent Surfaces (RISs) has been rising rapidly. This field has attracted continuous investment from global scientific research organizations and has strongly promoted the development of this field.

The hardware design of RISs is crucial for their application development. It not only determines the upper limit of their performance but also is directly related to the feasibility of practical deployment and cost-effectiveness. RISs are

Manuscript received **; revised **, 2024. This work is supported by the National Science Foundation (NSFC) for Distinguished Young Scholars of China (62225108), the National Key Research and Development Program of China (2023YFB3811502), the Jiangsu Province Frontier Leading Technology Basic Research Project (BK20212002), the Jiangsu Provincial Scientific Research Center of Applied Mathematics (BK20233002), the National Natural Science Foundation of China (62288101, 62201139, U22A2001), the Fundamental Research Funds for the Central Universities (2242022k60003, 2242024RCB0005), and the State Key Laboratory of Millimeter Waves, Southeast University (K202403 and K202316) (Corresponding author: Qiang Cheng).
Zhen Zhang is with the School of Electronic and Communication Engineering, Guangzhou University, China, and also with the State Key Laboratory of Millimeter Waves, Southeast University, Nanjing 210096, China (email: zhangzhen@gzhu.edu.cn).
Qiang Cheng are with the State Key Laboratory of Millimeter Wave, Southeast University, Nanjing 210096, China (email: qiangcheng@seu.edu.cn).

composed of two main parts: the reconfigurable array and the control module [15]-[18] (see Fig. 2). Each element within the array is embedded with tunable devices, such as positive-intrinsic-negative (PIN) diodes and varactor diodes. The response of the RIS element to wireless signals is altered by switching the biasing states of these tunable devices. The array control module is made up of digital signal processors, including microcontrollers (MCUs) and field-programmable gate arrays (FPGAs). These processors can switch the working states of active tunable devices in a programmable manner in real time, thus changing the responses of RISs to wireless signals. In the design of RISs, the structural design of the element poses a challenging problem. This is because it involves multiple design variables, such as the parameters of the dielectric substrate, the topological pattern of the element, and the working states of the tunable devices [19]-[21]. These design variables need to simultaneously satisfy several design objectives, including phase states, amplitude limitation, polarization mode, and the working bandwidth.

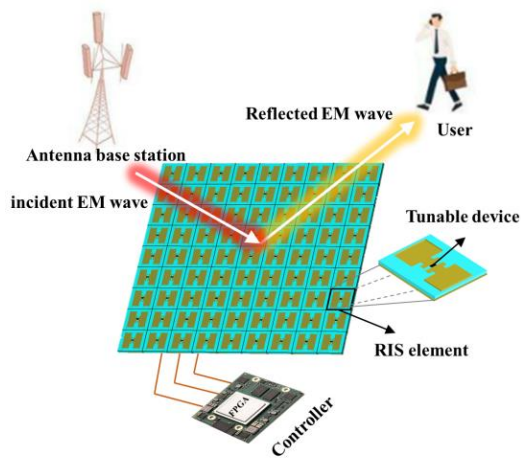

Fig. 2. Schematic diagram of the RIS that shows the scattering reconfiguration mechanism. The RIS element includes tunable components like diodes and varactors to manipulate the properties of the EM waves.

The mainstream design methods of RISs include: (a) the full–wave EM simulation method; (b) the machine-learning-based method; (c) the physical-model-based method. The following offers a brief introduction to the three methods:

(a) Full-Wave EM Simulation Method

This method uses full-wave EM simulation for parameter search (parameter sweeping) and parameter optimization of some classical structures [22]–[25]. For most designers, parameter search is an effective way to find the optimal combination of design parameters. However, when dealing with multi-parameter design problems, the search space becomes extremely large, increasing the complexity and challenge of the search process. To address the above issues, more intelligent search methods can be carried out with the aid of optimization algorithms. For example, M. F. Iskander et al. designed three-dimensional metamaterials using the genetic algorithm[22]. D. H. Werner et al. applied the adaptive ant colony algorithm to the design of three-dimensional frequency-selective surfaces [23]. Such methods show good universality, but exhibit the defect of poor efficiency since every change of the design variables requires an additional full-wave EM simulation. As a result, a large number of full-wave EM simulations are needed throughout the optimization process, resulting in a huge burden on computing resources and significant time consumption.

(b) Machine-Learning-Based Method

Its core idea is to construct a machine-learning model with low computational cost to replace the full-wave EM simulation, thus improving the efficiency of RIS design and shortening the whole design cycle [26]–[34]. J. M. Malof et al. designed all-dielectric metasurfaces using deep learning [26]. By combining the deep neural network and the fast-forward dictionary search algorithm, it enables the rapid identification of geometric parameters for desired transmission characteristics. For instance, T. Qiu et al. realized the automatic design of multi-bit intelligent reflective metasurfaces using deep learning [27]. To further reduce the number of samples required for machine-learning modeling, a physical knowledge inspired machine-learning method has been proposed [30]–[34]. The advantage of this method is that it can significantly decrease the time cost. However, it still relies on numerous full-wave EM simulations to establish the relationship model between design parameters and element responses. Thus, it is difficult to completely eliminate the dependence of machine-learning on full-wave EM simulations.

(c) Equivalent circuit-Based Method

The basic concept of this method is to convert the EM problem into a circuit problem under certain conditions, which employs well-known circuit theory to simplify the design process [35]–[40]. D. Abbott et al. modeled the band-pass frequency-selective surface with varactors by connecting the varactors in parallel with equivalent resistors, inductors, and capacitors [35]. F. Costa et al. established a circuit model by connecting the equivalent circuits of tunable devices and passive structures in parallel [36], [37]. The advantage of this method is that it can reduce the dependence on full-wave EM simulations. However, a significant limitation lies in the relatively low modeling accuracy for complex structures. This is because the equivalent circuits for certain multi-layer structures can be extremely intricate. There are numerous model parameters, and the parameter extraction process is quite complicated.

To address the above issues, a novel RIS design method based on multi-port network has been proposed. The principle of this method involves adding internal ports in the middle of the passive structure of RISs. Then, a relationship model is established between the internal port loads and the reflection coefficients of the RIS elements through the multi-port network. Moreover, by means of this relationship model, the reflection coefficients of the RIS elements corresponding to any internal port load can be quickly obtained.

In recent years, several works applying the physical model of multi-port network to the structural design of RISs have been published. For example, Q. Cheng, Z. Zhang, et al. used a two-port network to build a combined "field – circuit" model [41]. Based on this two-port model, the dependence of the tunable device parameters on the element reflection response can be rapidly evaluated without relying on full-wave EM simulation, and the effectiveness of the modeling method has been verified.

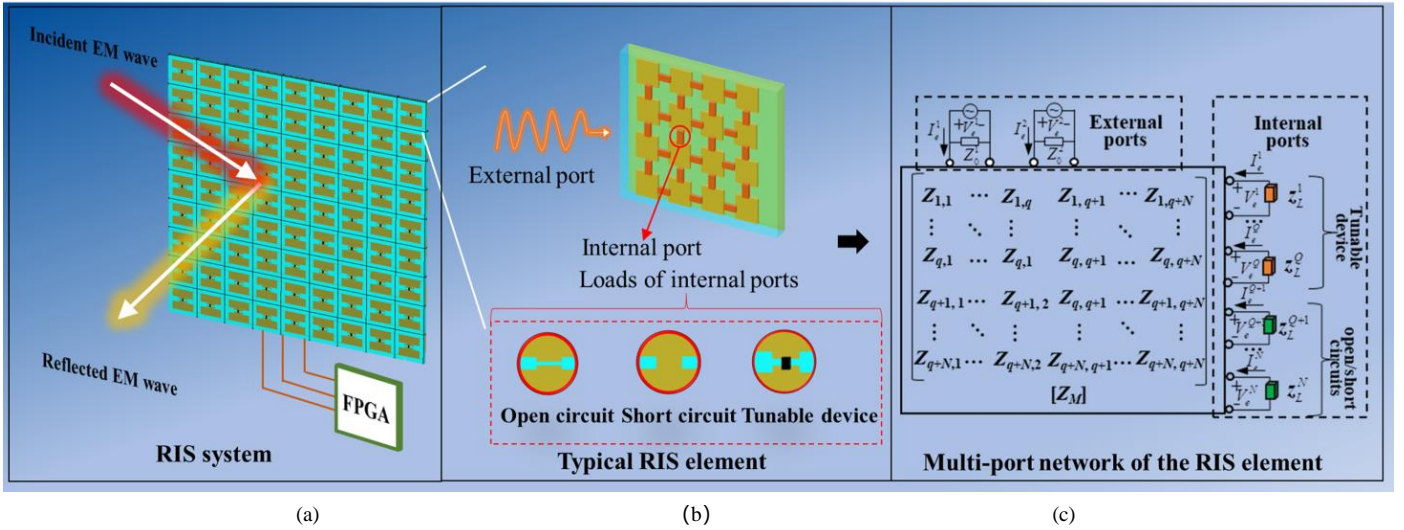

Fig. 3. Schematic of the RIS and the corresponding multi-port network model. (a) RIS system containing the reconfigurable array and the control module. (b) Typical RIS element which is discretized into multiple tiny patches connected by discrete internal ports, and each port can be connected with open circuit, short circuit or tunable device. (c) Multi-port network model of the RIS element.

On this basis, a macroscopic modeling method for RIS design based on multi-port network has been further developed [42]. In this method, the metallic pattern of the reconfigurable intelligent surface is discretized into multiple tiny patches, and multiple internal ports are added. A connection between multi-port loads and the EM response of the RIS is constructed using the multi-port network, and the structural parameters of the RIS are determined accordingly.

Furthermore, the incident EM waves with different polarizations can be regarded as sources at different external ports for the multi-port network. Thus, a new multi-port network model can be constructed to achieve dual-polarized RIS elements [43]. Additionally, to improve the extraction efficiency of the multi-port network, the residue-pole model can be employed to describe the dispersion of the multi-port network[44]. The above series of studies indicate that utilizing the multi-port network method enables the rapid design of RIS element and opens up a new path for improving the overall design efficiency. This work elaborates in detail the specific strategies for RIS element design using the multi-port network and provides a prospect for their future development.

## II. MULTI-PORT NETWORK MODEL OF RIS ELEMENTS

To achieve effective signal manipulation, RISs typically consist of hundreds to thousands of elements (Fig. 3(a)), as referenced in [45]-[48]. The process of RIS element design using a multi-port network involves discretizing each RIS element into multiple patches connected by discrete ports, as shown in Fig. 3(b). We refer to the port where the incident EM wave strikes as the external port, and the discrete ports between patches as internal ports. Impedance parameters for the multi-ports (including both internal and external ports) are extracted through EM simulations, and a relationship model is further established between the multi-port impedances, internal multi-port loads (such as open-circuit, short-circuit or tunable devices, etc.), and the reflection coefficient of the RIS element. With the initial structure remaining unchanged, this model allows for the rapid acquisition of the RIS element's reflection coefficient under arbitrary internal port loads, without the need for full-wave EM simulations. Furthermore, based on the type of loads connected to the internal ports, the internal ports are classified into passive internal ports and tunable internal multi-ports. Passive internal ports alter the topology of the passive metal patches of the RIS element by selecting either an open-circuit or short-circuit load, while tunable internal ports adjust the load impedance of tunable devices to change the operating states of the RIS elements, as depicted in Fig. 3(c). Finally, optimization algorithms are employed to obtain the optimal load impedance parameters for the internal ports, thereby finding the most suitable structure for the RIS element. In summary, designing RIS elements using a multi-port network primarily encompasses two aspects: RIS element modeling and performance optimization.

### A. RIS element modeling using the multi-port network

As shown in Fig. 3 (c), a ($N+q$)-port RIS element contains $q$ external ports (for EM wave excitation) and $N$ internal ports (for loads). According to [42], the equivalent impedance $Z_s(\omega)$ of the RIS element can be calculated by $V_e(\omega)$ (equivalent voltages at the external ports) and $I_e(\omega)$ (equivalent currents at the external ports), and as shown in Eq. (1).

$$\boldsymbol{Z}_s(\omega) = \frac{\boldsymbol{V}_0(\omega)}{\boldsymbol{I}_0(\omega)} = \boldsymbol{Z}_{qq}(\omega) - \boldsymbol{Z}_{q,q+N}(\omega) \cdot \left(\boldsymbol{Z}_L(\omega) + \boldsymbol{Z}_{q+N,q+N}(\omega)\right)^{-1} \cdot \boldsymbol{Z}_{q+N,q}(\omega) \quad (1)$$

In Eq. (1), $\omega$ is the angular frequency. $z_L^i(\omega)$ is the load impedance connected to the $i$th internal port. The load impedance matrix $\boldsymbol{Z}_L(\omega)$ is generated by diagonalizing the vector $z_L(\omega) = \left[z_L^1(\omega), \cdots, z_L^Q(\omega), z_L^{Q+1}(\omega), \cdots, z_L^N(\omega)\right]$. Namely, $Z_L^{i,i}(\omega) = z_L^i(\omega)$, $i, j=1,\ldots, N$, where the superscript of $Z_L^{i,i}(\omega)$ is the $i$-th row and $i$-th column element of $\boldsymbol{Z}_L(\omega)$. $\boldsymbol{Z}_{q,q}(\omega)$ is a $q \times q$ matrix that represents the self-impedance matrix of the external ports. $\boldsymbol{Z}_{q+N,q}(\omega)$ and $\boldsymbol{Z}_{q,q+N}(\omega)$ represent the impedance matrices that show the interactions among the internal and external ports $\boldsymbol{Z}_{q+N,q+N}(\omega)$ is a $N \times N$ matrix that represents the self-impedance matrix of the internal ports. $\boldsymbol{Z}_{q,q}(\omega)$, $\boldsymbol{Z}_{q,q+N}(\omega)$, $\boldsymbol{Z}_{q+N,q}(\omega)$, and, $\boldsymbol{Z}_{q+N,q+N}(\omega)$ form a $(q+N) \times (q+N)$ impedance matrix $\boldsymbol{Z}_M(\omega)$ (in Eq. (3))[42]. $\boldsymbol{Z}_0(\omega)$ is a $q \times q$ diagnal matrix, in which the elements on the diagonal are all equal to the impedance of free space.

$$\boldsymbol{Z}_M(\omega) = \begin{bmatrix} \boldsymbol{Z}_{q,q}(\omega) & \boldsymbol{Z}_{q,q+N}(\omega) \\ \boldsymbol{Z}_{q+N,q}(\omega) & \boldsymbol{Z}_{q+N,q+N}(\omega) \end{bmatrix}, \quad (2)$$

The relationship of the voltages and currents at the external and internal ports is shown in Eqs. (3-4)

$$\begin{bmatrix} \boldsymbol{V}_e(\omega) \\ \boldsymbol{V}_i(\omega) \end{bmatrix} = \begin{bmatrix} \boldsymbol{Z}_{q,q}(\omega) & \boldsymbol{Z}_{q,q+N}(\omega) \\ \boldsymbol{Z}_{q+N,q}(\omega) & \boldsymbol{Z}_{q+N,q+N}(\omega) \end{bmatrix} \cdot \begin{bmatrix} \boldsymbol{I}_e(\omega) \\ \boldsymbol{I}_i(\omega) \end{bmatrix}, \quad (3)$$

$$\boldsymbol{V}_i(\omega) = -\boldsymbol{Z}_L(\omega) \cdot \boldsymbol{I}_i(\omega), \quad (4)$$

where $\boldsymbol{V}_e(\omega)$ and $\boldsymbol{I}_e(\omega)$ are equivalent voltages and currents at the $q$ external ports (incident EM waves), respectively. $\boldsymbol{V}_i(\omega)$ and $\boldsymbol{I}_i(\omega)$ are the voltages and currents at the $N$ internal ports.

Once $\boldsymbol{Z}_s(\omega)$ is determined in Eq. (1), we can substitute it in Eq. (5) to achieve the reflection coefficient of the RIS element,

$$\Gamma(\omega) = \frac{Z_s(\omega) - Z_0}{Z_s(\omega) + Z_0}. \quad (5)$$

### B. Performance optimization of RIS elements

In the design of RIS elements, the core objective is to achieve efficient and flexible manipulation of EM wave properties, such as phase, amplitude, polarization, and frequency [49-50]. Taking the phase-modulated RIS as an example, we first utilize the aforementioned multi-port network model to analyze the reflection characteristics of the RIS element. Once the model is established, we proceed to define the design variable $x$ as the load of the internal ports and formulate the objective function for the reflection characteristics as follows:

$$G(\boldsymbol{x}, \omega) = W_1 \cdot \sum_{k=1}^{2^N} \left(|\varphi(\Gamma(\boldsymbol{x}, \omega)) - \theta^{(k)}|^2\right) - W_2 \cdot (\max(|\Gamma(\boldsymbol{x}, \omega)|_{\text{dB}}, A)). \quad (6)$$

Eq.(6) includes the amplitude and phase constraints for the $N$-bit RIS across all reflection states. In Eq. (6), $\theta(k)$ denotes the target phase for the $k$-th phase state of the RIS element, and $A$ denotes the amplitude constraint of the RIS element. $\varphi(\Gamma(\boldsymbol{x}, \omega))$ is the phase of the reflection coefficient, and $\max(|\Gamma(\boldsymbol{x}, \omega)|_{\text{dB}}$ is the maximum amplitude of the reflection coefficient in dB. $W_1$ and $W_2$ are the weights for the phase matching and the amplitude constraint of the reflection coefficient in the objective function, respectively. Both factors are typically set to 1. By optimizing the function $G(\boldsymbol{x}, \omega)$, the optimal design variable with the minimum function value is obtained. The optimization process is outlined as follows:

$$\boldsymbol{x}^* = \arg\min_{\boldsymbol{x}} \sum_{i=1}^{L} G(\boldsymbol{x}, \omega^i), \quad (7)$$

where $\omega^i$ is the $i$-th frequency point, $i=1, \ldots, L$, and $L$ is the number of frequency points.

### III. DEVELOPMENT PROCESS OF THE MULTI-PORT NETWORK DESIGN METHOD FOR RISs

In the past few years, the design of RIS elements using multi-port networks has undergone rapid development. The design approach was initially based on the macroscopic modeling of two-port network, and gradually turns to the modelling of multi-port network to obtain dual-polarized RIS structures with complex patterns. This section aims to provide an overview of the development of multi-port network design methods for RISs.

### A. Macroscopic modeling of two-port network

Fig. 4(a) presents a schematic diagram of the initial structure for an RIS element with two ports. The normally incident EM wave can be regarded as an excitation signal from an input port (Port $P$), also known as the external port. The port connected with the tunable device is a discrete port (Port $L$), also referred to as the internal port, as shown in Fig. 4(b) [41]. By setting the number of external ports and internal ports in Eq. (1) to 1, and simplifying it into a two-port network model, the relationship between the reflection coefficient of the RIS element and the impedance parameters of the tunable devices and the two-port network is obtained using Eq. (1) and Eq.(5). We express the relationship in the form of a function as follows::

$$\Gamma(\omega) = F\left(\boldsymbol{Z}_M^2(\omega), Z_L^a(\omega)\right) \quad (8)$$

where $\boldsymbol{Z}_M^2(\omega)$ represents the impedance matrix of the two-port network (in the form as presented in Eq. (2)), and $Z_L^a(\omega)$ denotes the impedance of the tunable device, which varies as the biasing voltage of the PIN diode or varactor changes.

Based on Eq. (8), we can obtain the reflection coefficient of the RIS element corresponding to different two-port impedance matrix $\mathbf{Z}_M^2(\omega)$ and load impedance matrix $Z_L^a(\omega)$ under various phase states. Since $Z_L^a(\omega)$ does not need to be obtained through EM simulations, we can analyze the EM response variations of the element when $Z_L^a(\omega)$ change without relying on full-wave EM simulations. It enables rapid selection and impedance setting of the tunable devices in the RIS element.

However, sometimes the phase range of the RIS element cannot be effectively enlarged by merely adjusting the impedances of the tunable devices. Therefore, in most cases, adjustment of both the passive structure and the impedance of the tunable devices are required to obtain an RIS element with expected performance.

### B. Macroscopic modeling of multi-port network

To increase the degree of freedom in designing the structure of RISs, the passive metal patch of the RIS element is discretized into multiple rectangular patches, with multiple internal ports added between these patches. The initial structure of the RIS element is depicted in Fig. 4(d), and its corresponding multi-port schematic is shown in Fig. 4(e) [42]. Through an EM simulation, the impedance parameters at all ports, including 17 internal ports and one external port, can be obtained. From Eq. (1), the relationship between the multi-port impedance, the internal port load impedance, and the reflection coefficient of the RIS element is obtained using Eq. (1) and Eq.(5). We express the reflection coefficient in the form of a function as follows:

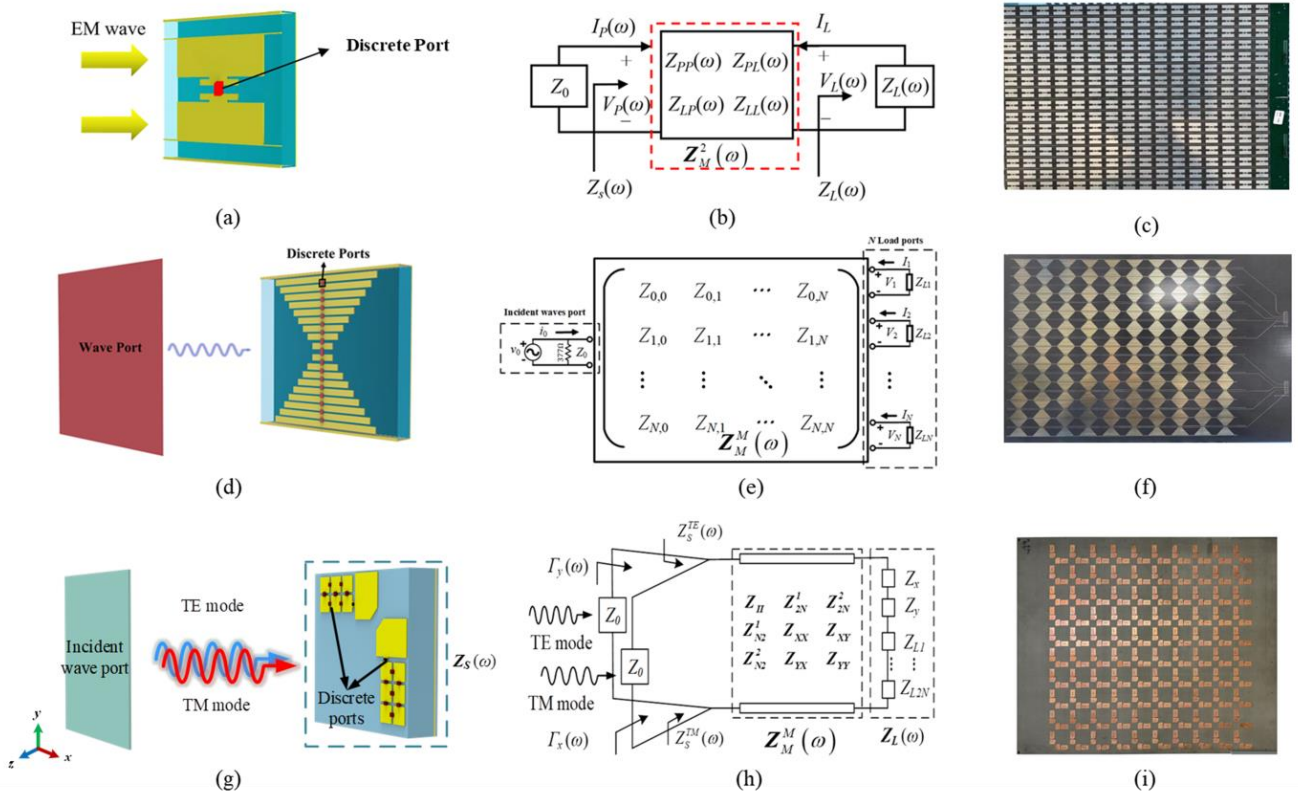

Fig. 4. (a-c) The initial structure, network model, and fabricated sample for a single polarized RIS based on the two-port network method [41]. (d-f) The initial structure, network model, and fabricated sample for a single polarized RIS based on the multi-port network method [42]. (g-i) The initial structure, network model, and fabricated sample for a dual- polarized RIS based on the multi-port network method [43].

Table I. The reported RIS designs using the multi-port network method

| Ref. | Number of Working States | Center Frequency | Polarization | Number of External Ports | Number of Internal Ports | Containing Tunable Device |
|---|---|---|---|---|---|---|
| [41] | 2 /8 | 9.6 GHz /3.3 GHz | Single polarization | 1 | 1 | Y |
| [42] | 8 | 2.6 GHz | Single polarization | 1 | 17 | Y |
| [43] | 2 | 5 GHz | Dual-polarization | 2 | 16 | Y |
| [51] | 16 | 2.4 GHz | Single polarization | 1 | 40 | Y |
| [52] | 16 | 2.4 GHz | Single polarization | 1 | 4 | Y |
| [53] | 10 | 2.4 GHz | Multi-polarization | 1 | 40 | Y |
| [54] | 8 /4 | 3.3 GHz /26.5 GHz | Single polarization | 2 | 1 | Y |
| [55] | 2 | 10 GHz and 14 GHz | Single polarization | 2 | 1 | Y |
| [56] | 2 | 207 GHz | Single polarization | 1 | 1 | Y |
| [44] | 8 | 3.3 GHz | Single polarization | 1 | 87 | Y |

$$\Gamma(\omega) = F\left(\mathbf{Z}_M^M(\omega), \left[\mathbf{Z}_L^a(\omega)\ \mathbf{Z}_L^p(\omega)\right]\right), \quad (9)$$

where $\mathbf{Z}_M^M(\omega)$ is the multi-port impedance matrix (in the form as presented in Eq. (2)), $\mathbf{Z}_L^a(\omega)$ is the load matrix of the tunable device, $\mathbf{Z}_L^p(\omega)$ is the load matrix of the passive structure.

Based on the multi-port model, appropriate load impedance at the internal port is selected using an optimization algorithm to achieve the desired EM response. In [42], leveraging this multi-port network model, a 3-bit RIS at 2.6 GHz was designed using a two-stage optimization algorithm. The fabricated RIS structure is shown in Fig. 4(f). Note that the RIS structure can work under single polarization.

*C. Dual-polarized RIS design with the multi-port network method*

To design dual-polarized RISs utilizing a multi-port network, the incident TE and TM EM waves on the RIS element are considered to be excited at two different external ports. The multi-port model is employed to rapidly calculate the reflected EM response, and a genetic algorithm is used to optimize the structural parameters in [43]. The initial structure of the RIS element is shown in Fig. 4(g), with the corresponding multi-port network depicted in Fig. 4(h). Furthermore, a low-cost multi-port network model for dual-polarized RIS design is established based on Eq. (1) and Eq. (5), which enables rapid calculation of the reflection coefficients of dual-polarized RIS elements (as shown in Eq. (11)). The genetic algorithm is then utilized to quickly and accurately select the internal port loads to optimize the RIS element, with the fabricated RIS shown in Fig. 4(i). The design time of this method is approximately 1/81 of that of traditional EM simulations [43].

$$\left[\Gamma_x(\omega)\ \Gamma_y(\omega)\right] = F\left(\mathbf{Z}_M^M(\omega), \left[\mathbf{Z}_L^a(\omega)\ \mathbf{Z}_L^p(\omega)\right]\right), \quad (10)$$

where $\mathbf{Z}_M^M(\omega)$ is the multi-port impedance matrix, $\mathbf{Z}_L^a(\omega)$ is the load matrix of tunable device, $\mathbf{Z}_L^p(\omega)$ is the load matrix of the passive structure. $\Gamma_x(\omega)$ is the reflection coefficient of the RIS in x-polarized direction, and $\Gamma_y(\omega)$ is the reflection coefficient of the RIS in y- polarized direction.

Recently, a growing number of literatures have focused on designing RIS structures based on multi-port network methods, as summarized in Table I. A compact RIS design based on the multi-port network was proposed to achieve 4-bit phase control at 2.4 GHz by meticulously selecting multi-port loads [51]. In [52], an active RIS with beam-scanning and amplification capabilities was designed using multi-port network, with the reflection gain of approximately 13 dB. In [53], an equivalent circuit model for RIS elements was established with the multi-port network for polarization RISs, and 10 polarization working states was obtained. In [54], [55], the multi-port network was utilized to expand the dataset for machine learning. In [56], the multi-port network was used to analyze the performance limit of RIS elements. In [44], the residual-pole model was utilized to accelerate the time cost of the multi-port network model for RIS elements.

## IV. FUTURE TRENDS

Currently, significant progress has been made in the design of RISs based on multi-port network methods; however, there are still challenges to further improve the design efficiency.

In the element design based on the multi-port network, the design efficiency is most profoundly influenced by the impedance extraction via full-wave EM simulations. The typical RIS element comprises four parts including tunable devices, metal patterns, dielectric substrates and metal ground, as depicted in Fig. 5 [54]. The multi-port impedance matrix must be re-extracted when any parameter in the four parts changes. In the meanwhile. The variation of the incident angle also has a notable effect on the reflection characteristics of the element. The re-extraction process will inevitably lead to the growth of the design time. This difficulty can be effectively resolved by further optimizing the following two models:

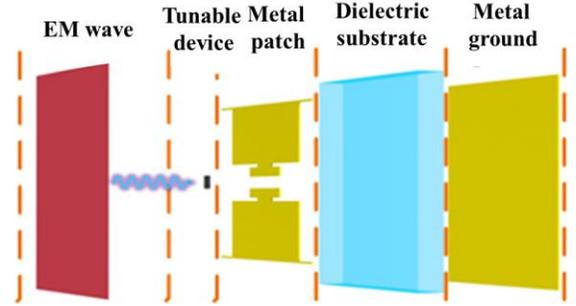

Fig. 5. Composition of a typical multi-layer RIS element [54].

(a) **Residual-Pole Model**: To enhance the efficiency of multi-port impedance extraction, the residual-pole model has been proposed to decrease the number of EM simulations required [44]. The zero-pole model is used to speed up obtaining $\mathbf{Z}$ parameters of the multiple ports, which significantly reduces the modeling time for the multi-port network model of RIS elements. However, this method does not fundamentally alleviate the impact of structural modifications and angle variations on the multi-port impedances. In the future, it is necessary to separate the influence of structural parameters and the incident angle on the element impedance, which will substantially improve the efficiency of multi-port impedance extraction.

(b) **Deep Learning Model**: Deep learning offers great potential to derive the relationship between different structural parameters and multi-port impedances, thereby improving the real-time nature of the optimization process, enabling faster and more responsive adjustments [57]. However, the primary challenge of deep learning is the need to establish a relationship between structural parameters and EM responses with numerous EM simulations as discussed before. However, this difficulty can be easily resolved by the multi - port method, since it can provide more low-cost datasets to accelerate the modeling speed of deep learning. Therefore, the combination of these two technologies is one of the key directions for future research.

On the other hand, in the future the multi-port network method can be extended to design the miniaturized and multi-functional RISs. By expanding the range of loads (see Fig. 6), we can not only design miniaturized RIS elements but also further explore the multi-functional integration of these elements. Additionally, we can combine RIS elements with antennas to achieve integrated scattering and radiation functionalities.

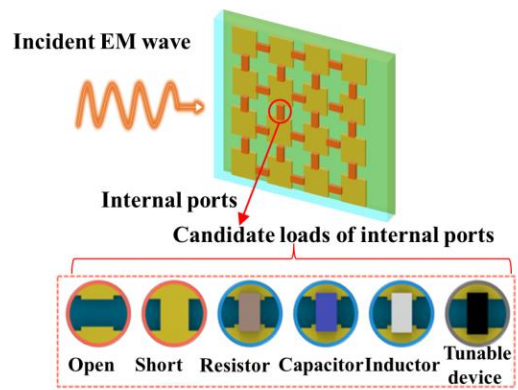

Fig. 6. Candidate loads connectable to internal ports in RIS elements

## V. Conclusions

In this paper, we provide an in-depth introduction to the design of Reconfigurable Intelligent Surfaces (RISs) based on multi-port network theory. We present a comprehensive discussion on the concept of multi-port network and its application in element design. A meticulous design procedure is described, and simultaneously, the associated challenges to further enhance the design efficiency are spotlighted. Moreover, we explore the future trends in this field to further improve the design efficiency.